\documentclass[11pt,twoside]{article}
\usepackage{newpasp,epsf}
\def\edcomment#1{\iffalse\marginpar{\raggedright\sl#1\/}\else\relax\fi}
\reversemarginpar

\begin{document}
\title{Current Problems for X-ray Emission from Radio Jets}
\author{D. E. Harris}
\affil{Smithsonian Astrophysical Observatory, MS-3 60 Garden
St. Cambridge, MA 02138 USA}

\begin{abstract}
A list is presented of known extragalactic radio jets which also have
associated X-ray emission.  The canonical emission processes for the
production of X-rays are reviewed and the sources are categorized on
the basis of our current understanding.  Although it seems clear that
the X-ray emission is non-thermal, the two competing processes,
synchrotron and inverse Compton emissions, arise from extremely high
energy (synchrotron) or extremely low energy (beaming models with IC
emission), relativistic electrons.  Only synchrotron self-Compton
emission from a few hotspots provides information on the `normal'
energy range of the electrons responsible for the observed radio
emission. 
\end{abstract}

\section{Introduction}

Until the advent of data from the Chandra X-ray Observatory, only a
few examples of X-ray emission from features in radio jets were
available.  Chandra now provides the required sensitivity and
resolution to distinguish knots and hotspots of radio galaxies and
quasars: we have already witnessed a doubling in the number of known
X-ray jet emitters and the number will surely continue to grow.

This paper focuses on the emission processes for the X-rays and is
meant to bridge the pre-Chandra epoch and where we are now.  Table 1
lists the sources known to me at this time.

\begin{table}
\caption{2000 July list of radio sources with jet related X-ray emission}
{\footnotesize
\begin{tabular}{lcclrccccl}
\tableline

Generic	&RA	&Dec	 & z	 &Dist.	&kpc/\arcsec   &Assoc. 	&Assoc. 	&PA  	&contact  \\
Name	&J2000	&J2000	&	&(H=50)  &(H=50)  &radio	&optical	&w.r.t. 	&  \\      
	&hh mm	&dd mm	&	 &(Mpc)  &        &     	&	&core	&        \\
\tableline

3C~120	&04 33	 &05 21	&0.0330	  &201	 &0.91 	&knot	 &no	&NW
&Harris, D.E. \\
3C~123	&04 37	 &29 40	&0.2177	 &1448	 &4.74	 &hs	 &no	 &110
&Hardcastle, M. \\
PictorA	&05 19	&\llap{-}45 46	&0.0350	 & 214	 &0.97	&W hs	 &yes	 &-80
&Roeser, H.-J. \\
PKS0637	&06 35	&\llap{-}75 16	&0.653	& 5197	 &9.22	&knots	 &yes	 &-90
&Schwartz, D.A. \\
PKS1127 &11 30	&\llap{-}14 49	&1.18	&11257	&11.48	&yes	  &?  	  &42
&Siemiginowska, A.L. \\
3C~273	&12 29	 &02 02	&0.1583	 &1025	 &3.70	&knots	&knots	 &190
&Marshall, H. \\
NGC4261	&12 19	 &05 49	&0.00737	   &44    &0.21   &yes	& no?	 &-90
&Birkinshaw, M. \\
M87	&12 30	 &12 23	&0.00427	  & 16	 &0.077  &knots	&knots	 &-60
&Biretta, J.A. \\
Cen A	&13 26	&\llap{-}42 49		&...  &  3.5  &0.017	 &?	 & ?	 & 70
&Kraft, R.P. \\
3C~295	&14 11	 &52 13	&0.45	& 3307	 &7.63	 &2hs	 &yes	&142/222
&Harris, D.E. \\
3C~371	&18 06	 &69 49	&0.510	 &3840	 &8.17	 &yes	 &yes&
&Sambruna, R. \\
3C~390.3 &18 42	 &79 46	&0.0561	  &346	 &1.50	 &hsB	 &hsB	&-10
&Harris, D.E. \\
Cyg A	&19 59	 &40 44	&0.0560	  &345	 &1.50	 &2hs	 &no
&110/280	&Harris, D.E. \\
\tableline
\tableline
\end{tabular}
}

\normalsize

Notes and Comments

q$_0$ = 0

Morphology: we generally use 'linear' to mean a smooth feature
(e.g. the jet in Pictor A); 'knot' as a distinct brighter feature in a
jet that continues past the feature; and 'hotspot' (`hs') either as
the terminal bright enhancement at the end of an FRII jet, or as one
of the multiple features associated with the termination of a jet.
Normally 'knots' are found on the inner portion of FRI jets whereas
'hotspots' are mostly at the ends of FRII jets.  However, we are not
trying to impose distinct definitions, and infer no physical
differences beyond these generalities.

The names appearing in the last column are meant to be a guide when you 
want to email someone to find out more; they are obviously not meant to
provide all the references.  As more Chandra and XMM data are published,
these names will be updated (eventually on a website).

\end{table}

\section{The emission processes}

Although thermal models were considered with the initial detections of
jets, in all cases two rather serious problems are encountered: (1)
how can we explain the existence of over-pressured hot gas which is
both far from the galactic nucleus and has no obvious source of
confinement to maintain the integrity of the emitting region, and (2)
why are the Faraday rotations and depolarizations predicted for even
quite modest (10 $\mu$G) magnetic fields absent?  These arguments have
been used against thermal models for the hotspots of Cygnus A (Harris,
Carilli, \& Perley 1994), the 25\arcsec\ knot of 3C~120 (Harris et
al. 1999), and many other X-ray emitting hotspots and knots.

Synchrotron emission could be considered the `process of choice' for
X-rays from knots in radio jets mainly because the optical
polarization observed in sources such as M87 is convincing evidence
that the optical emission as well as the radio emission comes from the
synchrotron process.  Demonstrations that the X-ray intensity was
consistent either with a single power law extrapolation from radio and
optical bands (e.g. 3C~390.3, Harris, Leighly, \& Leahy 1998) or with
a broken power law (e.g. knot A in M87, Biretta, Stern, \& Harris
1991) were taken as circumstantial evidence that the X-rays were also
generated by synchrotron emission.  Required for this model is the
presence of electrons with Lorentz factor $\gamma > 10^7$ (cf.
values of 10$^5$ for optical emission).  In the typical equipartition
fields of B $\approx 10^{-4}$G, the radiation half-life of the X-ray
emitting electrons would be of order 10 years.  No overriding
theoretical (or other) objections preclude this sort of model.

Inverse Compton emission has a distinct advantage over X-ray
synchrotron emission in that the extremely high energy for the
emitting electrons is not required.  What is required are both enough
electrons and enough energy density in photons of the proper energy to
produce the desired scattered photons:
$\nu_{out}~\approx~\nu_{in}~\times~\gamma^2$.

Although synchrotron self-Compton (SSC) emission is mandatory for all
synchrotron sources, it is usually the case that the photon energy
density ($u_{\nu}$) is significantly smaller (i.e. by factors of 100
or more) than the energy density of the magnetic field ($u_B$) and
thus the major energy loss for all electrons is via synchrotron
emission (and the SSC component is too weak to observe or distinguish
from other emissions).  Thus we have the often quoted result that the
ratio of IC to synchrotron losses is given as:\\

\hspace{0.8 in}R~=~$(dE/dt)_{ic}/(dE/dt)_{sync}~\approx~u_{\nu}/u_B~\approx~L_{ic}/L_{sync}$.\\

Another advantage of SSC over other processes is that accurate
predictions of source strength can be made.  If you have a reasonable
estimate for the emitting volume, you can calculate the photon energy
density from the synchrotron spectrum and then figure out what B field
you need to have the right number of electrons to give the observed
(radio) synchrotron and the observed (X-ray) IC emissions.

\section{Sources which 'work' with SSC}

There are three FRII radio galaxies with convincing SSC X-ray emission from
their (terminal) hotspots: CygA (Harris et al. 1994); 3C~295 (Harris
et al. 2000); and 3C~123 (Hardcastle, Birkinshaw, \& Worral 2000).
For all three of these sources, predicted SSC emission was calculated
from the radio data, proposals were written, and the hotspots were
detected as predicted.  In all cases, the average magnetic field
strengths derived from the SSC equations are consistent with
equipartition fields for the case of little or no contribution to the 
particle energy density from relativistic protons.

The hotspots of these sources are, so far, the only resolved
structures for which convincing SSC models have been published.  Since
the `Proton Induced Cascade' (PIC) model is a sort of a `sister process'
to SSC (it requires extremely energetic relativistic protons and a
high photon energy density; Mannheim, Kr\"{u}lls, and Biermann 1991) we
have the following `options' for these hotspots:

\begin{center}
\begin{tabular}{lll}
    CASE A              &             CASE B            &    CASE C \\
B $\approx$ 200 $\mu$G  &        B $>>$ 200 $\mu$G      & B $>>$ 200$\mu$G \\
B $\approx$ B$_{eq}$    &        B $>>$ B$_{eq}$        & B $\approx$B$_{eq}$ \\
k=0 or 1                &        k=0 or 1               & k $>>$ 1 \\
X-ray from SSC          &        X-ray from ?           & X-ray from PIC/synchrotron
\end{tabular}
\end{center}

\section{Sources which 'work' with synchrotron emission}

Synchrotron emission has classically been thought of as an extension
of the radio/optical spectrum, as in the case of M87 where it was
already evident 15 years ago that thermal models had problems, and the
jet emissions were believed to be synchrotron on the basis of the
observed polarization.

\begin{itemize}

\item{M87 knot A (and D and B)}

The synchrotron parameters were derived from Einstein data (Biretta et
al. 1991) for knots A, D, and B.  Wilson now has Chandra data, and it
remains to be seen how the argument will go for the other knots
(Marshall, private communication has the zero order image from an
HETG observation which shows that most of the radio/optical knots are
detected).

\item{3C~390.3 - Hotspot B}

This source exhibits rather peculiar circumstances: a hotspot where
the northern jet appears to collide with a dwarf galaxy in the
3C~390.3 group.  Prieto has published optical photometry (Prieto \&
Kotilainen 1997) and the spectrum is a fairly good approximation to a
single (or broken) power law.  The group at Heidelberg is working on a
Chandra observation of this source, but the results are not yet
public.

\item{3C~273 knots A1 and B1}

From archival HST data and Chandra observations, all of the prominent
optical knots in this jet are seen to have X-ray counterparts, see
fig. 1.  Three of these are bright enough and
separated enough to derive reasonably accurate X-ray photometry.  Knot
A1 has essentially a straight power law spectrum from the radio to the
X-ray, with $\alpha \approx$ 0.8.  B1 can also be fit with a broken
power law, whereas D/H3 belongs to the classification of the next
section.

\end{itemize}

\begin{figure}
\plotone{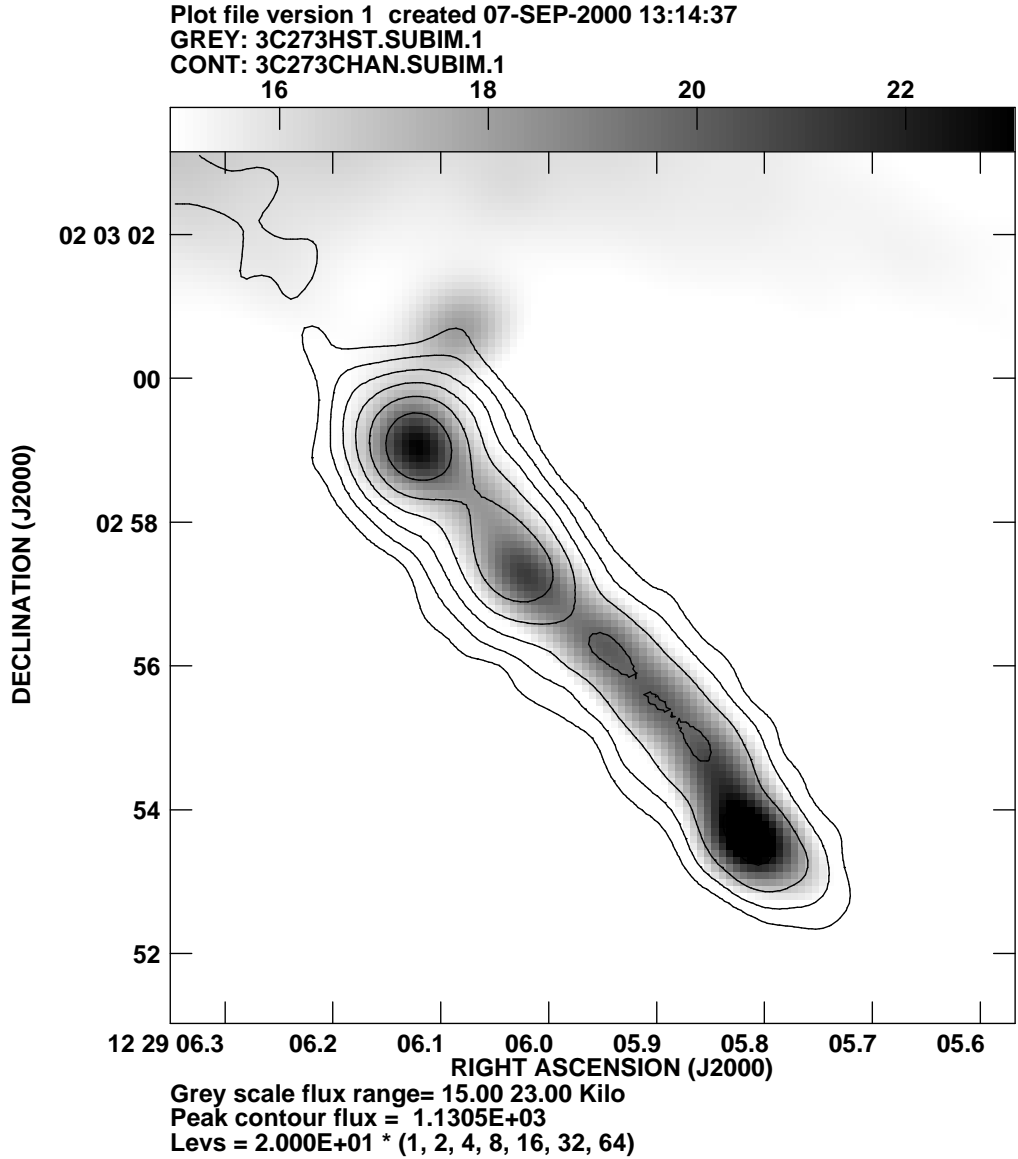}
\caption{An X-ray/Optical overlay of the 3C 273 Jet.  The contours are
from Chandra data which have been slightly smoothed.  Contour levels
increase by factors of two.  The grey scale is from the HST, smoothed
with a Gaussian to approximate the resolution of the X-ray data.
Further images can be found in Marshall et al. 2000.}
\end{figure}

\section{Sources with problems}

Historically, jet X-ray targets were selected almost solely on the
grounds that a radio feature had been detected in the optical, thereby
indicating that perhaps the electron spectrum was not cut off at the
usual value for most radio sources.  Actual detections were often
serendipitous since there had been no expectation that the detected
features would be X-ray emitters.  Two obvious examples of this are 3C
120 and PKS0637.

For 3C~120, we proposed ROSAT observations in an attempt to detect the
inner jet seen in the optical.  Instead, we found emission form a
relatively unremarkable radio knot further out in the jet (Harris et
al. 1999).  Since a good upper limit in the optical precluded a
monolithic single or double power law fit from the radio to the X-ray,
none of the simple emission models was acceptable.

In the case of the X-ray jet in PKS 0637 (Schwartz et al. 2000 \&
Chartas et al. 2000), the quasar was targeted by Chandra for focusing
tests on the assumption that it would be a point source.  Although
there is optical emission from this jet, the intensities are low
enough that the situation is similar to 3C~120.

Other examples for which the spectrum appears to have a cutoff in the
optical are knot D/H3 in the 3C~273 jet and the hotspot of Pictor A
(R\"{o}ser, private communication and Wilson, Young, \& Shopbell, 2001).

For these sources, the SSC calculations yield predictions which are 2
orders of magnitude or more below the observed intensities and the
spectrum shows that extrapolation of the radio/optical flux densities
would also under predict the X-ray intensity by a large factor.

\section{Conclusions}

\subsection{Current status}

From the observations, we have a few hints to guide us:
	
\begin{itemize}

\item{With a few exceptions (e.g. smooth, featureless jets such as
that emanating from the core of Pictor A, Wilson et al. 2001; and the
3C~273 jet where the radio emission from the optical/X-ray knots is
swamped by brighter, larger structures) we usually see large gradients
in the radio surface brightness which implies that strong shocks are
present at the sites of X-ray emission.}

\item{For the data I have seen, \em all\em\ of the `problem' X-ray
knots occur only on one side of the core.  This suggests that
relativistic beaming may be present.}

\end{itemize}

\subsection{What are the options for the `problem' features?}

One possibility is the presence of ad hoc flat spectrum components
embedded in the larger scale structures responsible for the radio and
optical emissions.  We use the term `ad hoc' in the sense that such a
distinct spectral component has not been detected at other wavebands.
Can we devise models with flat spectra so that the optical emission is
below current limits and the radio components are too faint to be
detected or are lost in the brighter emission from steeper spectrum
components?  This sort of model was suggested for the 25\arcsec\ knot
in 3C~120 (Harris et al. 1999), but now we require an extension to
distributed features (e.g. the narrow jet from Pictor A or the
underlying 'smooth' emission in 3C~273).  Current limits call for the
X-ray to optical spectral index to be flatter than 0.5.  Although
canonical shock acceleration provides only for $\alpha >$ 0.5, oblique
shocks can produce flatter spectra (Gieseler and Jones 2000).

The other viable explanation is based on the premise that even far
from the core, the jet is moving with significant relativistic bulk
motion, $\Gamma$ (Celotti, Ghisellini, \& Chiaberge 2000; Tavecchio et
al. 2000).  In this scenario, quite normal radio knots created with
the usually invoked shocks in the jet experience an energy density of
the CMBG which is $\Gamma^2$ larger than for a source at rest.  When
both the radio and inverse Compton emissions are beamed towards the
observer, it is possible to produce the observed intensities with
beaming factors of order 10.  This model is appealing in that it would
solve a long standing problem, but it remains to be seen if it is not
liable to `over-production': can it be demonstrated that only a select
few of all radio jets will fulfill the required conditions?  The
optical morphology of the 3C~273 jet bears a striking resemblance to a
barber shop pole: regularly spaced segments of a helix appear to be
the cause of the knots, see figure 2.  If most jets consist of
particles moving along helical paths, beaming can become operative
even if the overall jet direction does not make a small angle to the
line of sight.

\begin{figure}
\plotone{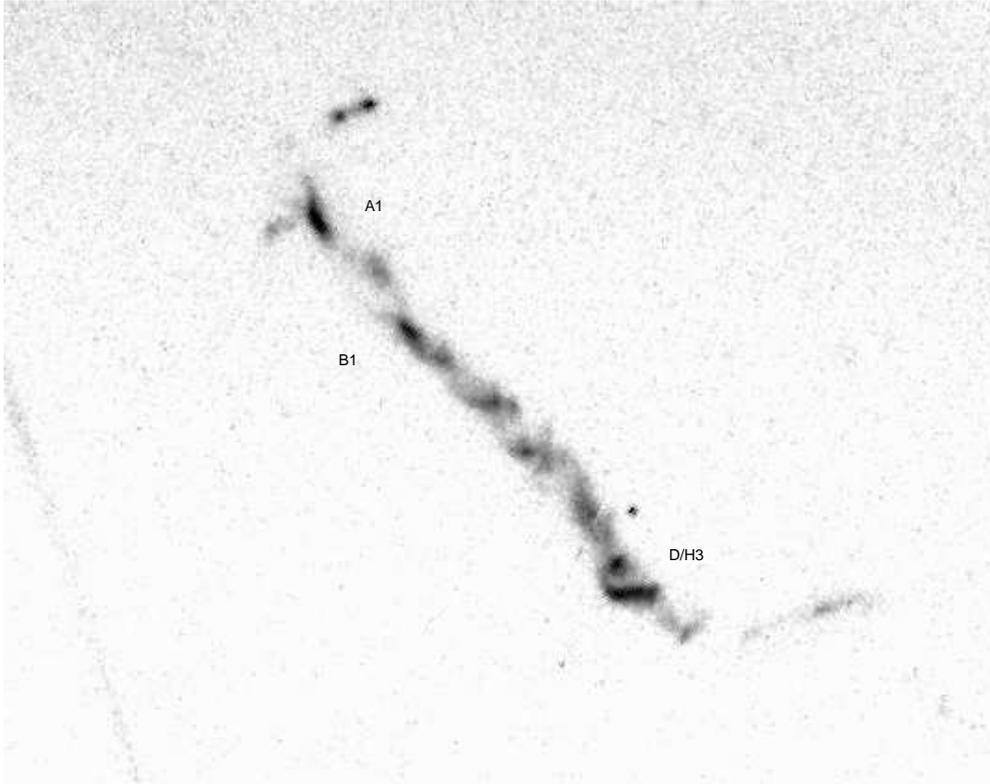}
\caption{An archival HST image of the 3C 273 jet.}
\end{figure}

\subsection{What does this mean for particles and fields in RG?}

	For SSC from terminal hotspots: B(ssc) $\approx$ B(eq) with no
protons.  The inference is that FRII jets consist of e+/e- and that
entrainment is not significant.  The SSC model also provides some
reassurance that conventionally computed equipartition fields stand a
good chance of being correct.  NB: PIC with higher B field is an
alternate model and would most likely require a normal plasma for the 
jet fluid.

	For synchrotron models of knots: Unless bulk velocities are
playing an important role via boosting, $\gamma \ga 10^7$
and halflives of order a few years are required.  In the beaming
model, the X-ray emission originates from $\gamma$ of order a few
hundred and we would be evaluating the electron spectrum at the very
lowest energies.

\acknowledgments

Many colleagues made significant contributions to the results reported
here.  Although it is impossible to name them all, a few are:
C. Walker who worked on our new VLA data of 3C~120, J. P. Leahy who
did the same for 3C~390.3, J. Grimes produced the results for the
optical data on 3C~273, H. Marshall is leading the CXC work on 3C~273
and L. David, P. Nulsen, and T. Ponman obtained the Chandra spectral
results on 3C~295.  Also I thank H. Krawczynski and W. Tucker for
extensive discussions on jets and their beaming properties.  This work
was partially supported by NASA contracts NAS5-99002 and NAS8-39073.

\end{document}